# Ultrafast µeV-Precision Bandgap Engineering in Low-Dimensional Topological Insulators

Peng Tan[1]†, Yuantao Chen[1]†, Yuqi Zhang[2,9,10], Hanyan Cheng[1], Guoyu Xian[3,8], Ming Cheng[4], Minghong Sun[1], Jiaxin Yin[1,5,6], Feifan Wang[3,8], Yaxian Wang[3], Yanjun Liu[4], Mingyuan Huang[1,6], Zhiwei Wang[2,9,10], Yugui Yao[2,9,10], Sheng Meng[3,7,8], Li Huang[1,5,6]*, Yanan Dai[1,5,6]*

[1]*Department of Physics, State Key Laboratory of Quantum Functional Materials, and Guangdong Basic Research Center of Excellence for Quantum Science, Southern University of Science and Technology, Shenzhen 518055, China.*

[2]*Key Laboratory of Advanced Optoelectronic Quantum Architecture and Measurement, Ministry of Education, School of Physics, Beijing Institute of Technology, Beijing 100081, China.*

[3]*Beijing National Laboratory for Condensed Matter Physics and Institute of Physics, Chinese Academy of Sciences, Beijing 100190, China.*

[4]*Department of Electrical and Electronic Engineering, Southern University of Science and Technology, Shenzhen, Guangdong, 518055 China.*

[5]*Quantum Science Center of Guangdong–Hong Kong–Macao Greater Bay Area (Guangdong), Shenzhen 518045, China.*

[6]*Guangdong Provincial Key Laboratory of Advanced Thermoelectric Materials and Device Physics, Southern University of Science and Technology, Shenzhen 518055, China.*

[7]*School of Physical Sciences, University of Chinese Academy of Sciences, Beijing 100190, China.*

[8]*Songshan Lake Materials Laboratory, Dongguan, Guangdong 523808, China*

[9]*Beijing Key Lab of Nanophotonics and Ultrafine Optoelectronic Systems, Beijing Institute of Technology, Beijing 100081, China.*

[10]*International Center for Quantum Materials, Beijing Institute of Technology, Zhuhai 319000, China.*

*Contact author: daiyn@sustech.edu.cn; huangl@sustech.edu.cn

†These authors contributed equally to this work

## Abstract:

Precise and ultrafast control of electronic band structures is a central challenge for advancing quantum functional materials and devices. Conventional approaches—such as chemical doping, lattice strain, or external gating—offer robust stability but remain confined to the quasi-static regime, far from the intrinsic femto- to picosecond dynamics that govern many-body interactions. Here, using cryogenic transient reflectance spectroscopy, we realize dynamic bandgap engineering in the anisotropic topological insulator α-$Bi_4Br_4$ with unprecedented micro–electron-volt (µeV) precision. The exceptional sensitivity arises from the cooperative action of long-lived topological carriers, stabilized by restricted bulk-to-edge scattering phase space, together with symmetry-resolved coherent phonons that modulate inter-chain hopping. These channels jointly modify Coulomb screening and interband transitions, enabling both gradual and oscillatory control of the electronic structure. Supported by first-principles and tight-binding theory, we further demonstrate a dual-pump coherent control strategy for continuous, mode-selective tuning of electronic energies with µeV accuracy. This framework paves the way for ultrafast on-demand band-structure engineering, pointing toward new frontiers in quantum optoelectronics, precision measurement in molecular and biological systems, and attosecond control of matter.

**Introduction**

The discovery of exotic quantum states of matter often hinges on the ability to finely tune electronic structures and the Fermi level in solids. Paradigm-setting discoveries such as the quantum Hall effect and dissipationless surface transport demonstrate how band topology can be reshaped by external fields [1-3]. More recently, moiré superlattices and atomically thin transition-metal dichalcogenides have shown that twisting, straining, or dielectric engineering can generate flat bands, correlated insulators, and tunable excitonic gaps [4-9]. While powerful under equilibrium conditions, they remain restricted to the DC limit and not designed to capture the intrinsic dynamics of electrons and lattices that govern quantum-state formation. Overcoming this limitation calls for ultrafast approaches that operate on the natural timescales of many-body interactions and deliver band-structure control with a resolution far beyond static means.

Ultrafast light pulses provide a powerful route beyond static control, enabling electronic bands to be manipulated on the natural timescales of electron, phonon, and quasiparticle interactions. In monolayer transition-metal dichalcogenides, for example, photoinjected carriers dramatically alter dielectric screening, producing giant excitonic shifts of hundreds of meV within femtoseconds [10, 11]. Strong light fields can further drive coherent reshaping of electronic dispersions: Floquet engineering with mid-infrared or terahertz pulses periodically dresses electronic states, generating replica bands and dynamical gaps [12, 13]. Extending these concepts to the lattice, coherent excitation of phonon modes has been predicted to tune bands through dynamical polarization or phonon-assisted Floquet mechanisms [14-16], and has been demonstrated to modulate charge-density-wave and superconducting gaps in correlated materials [17-19], as well as excitonic and interband resonances in semiconductors [20-22]. Despite these advances, resolving the intertwined influence of carrier screening and lattice dynamics on out-of-equilibrium band structures remains a key challenge, reflecting the fundamental complexity of electron–phonon interactions [23].

An ideal platform to investigate the cooperative roles of electronic and lattice interactions in band-structure control is the quasi-1D topological insulator α-$Bi_4Br_4$. This material hosts a sizable bulk bandgap together with topologically protected edge states that sustain the quantum spin Hall effect at room temperature [24-26]. Such properties point to potentially long-lived photocarriers stabilized against rapid relaxation, leading to strong Coulomb screening [27-29]. Its quasi-1D crystal structure, characterized by highly anisotropic electronic bands, could further enable momentum-and polarization-selective manipulation of carrier dynamics [30], while the heavy bismuth and bromine constituents impart strong lattice polarizability that enhances coherent phonon responses to optical excitation [31, 32]. These combined attributes position α-Bi$_4$Br$_4$ as a model system in which electronic screening and lattice dynamics can be disentangled and harnessed, establishing a versatile framework for ultrafast band engineering in low-dimensional quantum materials.

Here we use transient reflectance spectroscopy to reveal that femtosecond near-infrared excitation in α-Bi$_4$Br$_4$ induces a nonlinear redshift of optical resonances by up to ~25 meV in the visible spectrum—an effect far exceeding that observed in conventional bulk semiconductors. Superimposed on this carrier-driven response, distinct coherent phonon modes periodically modulate the same resonances with amplitudes approaching ~1 meV, showing a finely resolved lattice-driven control of the electronic structure. *Ab initio* calculations combined with a tight-binding model of the zigzag Bi chain [33] identify the key roles of quasi-1D charge confinement, topologically protected relaxation pathways, and phonon-induced polarization dynamics in shaping the nonequilibrium band structure. Finally, by implementing a phase-controlled dual-pump scheme, we achieve μeV-level tuning of band energies in a mode-selective manner. Together, these results provide a general strategy for light-driven band-structure engineering in low-dimensional quantum materials, uniting incoherent carrier screening with coherent lattice dynamics to achieve ultrafast control at unprecedented precision [16, 34-36].

**Results and Discussion**

Figure 1a illustrates the crystal structure of α-Bi$_4$Br$_4$ and the mechanisms of carrier- and phonon-driven band engineering explored here. This monoclinic lattice (space group C2/m) consists of quintuple Bi–Br layers stacked along *c*, forming infinite molecular chains along *b* that repeat periodically along *a*. This chain-like motif cleaves into needle-shaped flakes aligned with *b*, enabling quasi-1D carrier localization (yellow sphere in Fig. 1a) and giving rise to pronounced electronic and optical anisotropy. Such anisotropy is quantified by $\rho_S = (S_b - S_a)/(S_a + S_b)$, where $S_{a,b}$ denotes the absorption (*α*), reflectance (*R*), and transmission (*T*) measured with *a*- or *b*-polarized light (see Fig. S1). Figure 1b presents the anisotropy spectra from 1.7 to 2.1 eV: while transmission and absorption ($\rho_T$, $\rho_\alpha$) are positive throughout, reflectance ($\rho_R$) is negative with a ~20% minimum near 2 eV. These results highlight the strong polarization dependence of optical transitions in α-Bi$_4$Br$_4$, particularly in the visible regime where the anisotropy is most pronounced.

Beyond its electronic anisotropy, α-Bi$_4$Br$_4$ exhibits strongly polarization-dependent phononic responses that shape its dielectric environment, as shown by the polarized Raman spectra in the parallel configuration (Fig, 1c, *a*- and *b*-axes at 0° and 90°; cross-polarized spectra in Fig. S2). Owing to inversion symmetry, only A$_g$ and B$_g$ modes are observed, with more than ten Raman peaks detected due to the large unit cell.

The polarization-integrated spectrum (white curve in Fig. 1c) allows quantitative mode assignment, including the 77 cm$^{-1}$ and 115 cm$^{-1}$ fingerprints of the α-phase of Bi$_4$Br$_4$ [37, 38]. Here we focus on two low-frequency modes at 22.7 cm$^{-1}$ and 42.0 cm$^{-1}$, whose polarization-dependent intensities (Fig. 1d) show the twofold symmetry of A$_g$ phonons in the C2/m point group. These modes correspond to breathing-like vibrations of Br atoms perpendicular to the quasi-1D chains (Fig. 1a). Combined with the anisotropic charge distribution revealed by optical spectroscopy, such phonons provide key channels for dynamic modulation of the band structure.

To probe light-induced band modification, we first measured polarization- and fluence-dependent transient reflectance (ΔR/R) at pump–probe overlap (Δt = 0 ps), as shown in Figs. 2a,b. The pump polarization had negligible effect on the spectra (Fig. 3), consistent with rapid momentum relaxation that redistributes carriers across the Brillouin zone. By contrast, the probe polarization revealed striking anisotropy: pronounced resonances emerged near ~1.9 eV for *b*-polarized detection, while *a*-polarized signals were nearly an order of magnitude weaker. For *b*-polarization, both the magnitude and spectral weight of the resonances increased nonlinearly with carrier density $n_e$, evolving into Lorentzian-like features superimposed on a broad incoherent background (Fig. S4). The dominant resonance exhibits a fluence-dependent redshift of up to ~25 meV (Fig. 2c), following a power-law dependence with a scaling factor of ~2. This indicates a substantial gap reduction scaling with the interparticle spacing ($r_s$) as $\Delta E \sim r_s^{-6}$ that is steeper than the $\Delta E \sim r_s^{-4}$ behavior in typical bulk semiconductors [39]. This enhanced nonlinearity originates from quasi-1D confinement and anisotropic screening unique to α-Bi$_4$Br$_4$ [40]. Burstein–Moss band filling may introduce a partial blueshift [41, 42], implying that the intrinsic gap reduction is even larger. Overall, these results demonstrate a pronounced light-induced narrowing of the optical gap, highlighting strong many-body interactions and anisotropic carrier accumulation that dynamically reshape the band structure.

As the pump–probe delay Δt is scanned, the carrier-induced gap narrowing recovers on a few-picosecond timescale, as seen in the transient spectra in Fig. 2d. To track the resonance shifts, we evaluate the spectral derivative $R_\Delta(\omega) = \partial(\Delta R/R)/\partial\omega$ based on the raw data in Figs. S5a,b. The zero-crossing points $R_{\Delta,0}(t)$ reveal not only an exponential-like recovery of the gap but also pronounced oscillations with amplitudes of ~1 meV, evidencing coherent, mode-selective tuning of the electronic structure consistent with molecular dynamics predictions[16]. Fourier analysis of $R_{\Delta,0}(t)$ identifies two dominant frequencies at ~0.68 THz and ~1.26 THz, coinciding with both steady-state Raman modes (22.6 cm$^{-1}$ and 42.0 cm$^{-1}$ modes in Fig. 1c) and the coherent phonon modes resolved directly in ΔR/R (Fig. 2e). The corresponding energy modulations are ~1.44 meV and 1.12 meV, respectively (see Methods for details).

To elucidate the mechanisms of band modification, we analyzed the probe-energy dependence of transient reflectivity for *b*-polarized detection, which is most sensitive to the optical gap renormalization (*a*-polarized results are shown in the Supplementary Materials). The relaxation dynamics change markedly across the spectrum: above 1.7 eV the signal follows a nearly single exponential decay with an initial reduction in reflectivity, whereas at higher energies it evolves into a tri-exponential response with a sub-100 fs sign reversal to positive. These behaviors can be attributed to three competing processes: (i) excited-state absorption (ESA), (ii) band filling (BF), and (iii) electron–hole (e–h) recombination. Auger heating can be excluded, as it would yield a nonlinear increase in carrier density [43]. Again, coherent phonon excitations are observed across the spectrum. To capture both contributions, we employ a composite model consisting of tri-exponential decay terms for the electronic processes and damped sinusoidal terms for the phonons. Examples of the fitted curves are shown in Fig. S6, and Fig. 3b summarizes the extracted lifetimes of the ESA and BF processes.

The ESA contribution is confined to low probe energies, with a lifetime of ~8.5 ps at 1.4 eV that decreases to ~4.1 ps at 1.7 eV, reflecting the prolonged persistence of hot carriers near the Fermi level where accumulation is enhanced. Above ~1.8 eV, the sign reversal of $\Delta R/R$ arises from BF, as state occupation in the conduction band blocks ground-to-excited-state transitions. Band filling dynamics largely mirror those of ESA, both being governed by the available conduction-band density of states. Their lifetimes decrease with probe energy below ~1.7 eV; however, unlike ESA, the BF lifetime shows a slight upturn at higher energies, likely due to the resonant transition near 2 eV.

At longer delays, excited carriers recombine and the system relaxes toward equilibrium on a hundreds-of-picoseconds timescale. Such long-lived carriers are characteristic of topological insulators: electrons in bulk parabolic bands face restricted relaxation into Dirac-like edge states due to their low density of states. [27, 29]. This phase-space bottleneck is reinforced by the limited availability of phonon modes with suitable energy and momentum. A direct consequence is the buildup of a surface photovoltage (SPV), which generates an internal electrostatic gradient between bulk and surface and alters the surface work function [28]. Evidence of this effect is shown in Fig. S7, where momentum-integrated time-resolved photoemission from thin α-$Bi_4Br_4$ flakes reveals a transient downward Fermi-level shift that recovers on the tens-of-ps scale, consistent with SPV dynamics in TIs. By contrast, the phonon lifetimes $\tau_{ph}$ remain essentially energy independent, in agreement with the Raman nature of coherent phonon generation [44, 45].

The ESA and BF processes, which directly determine carrier lifetimes, are strongly correlated with excitation density. Their competition in shaping the transient reflectivity is most evident on the low-energy side of the spectrum (Fig. 3c, 1.5 eV probe), where the positive hump in $\Delta R/R$ shifts to longer delays with increasing $n_e$. The corresponding relaxation times extracted for ESA and BF are summarized in Fig. 3d.

For BF, the lifetime increases from ~4.8 ps to ~6.3 ps, consistent with a phonon bottleneck that prolongs carrier relaxation [46, 47]—a hallmark of topological insulators [27-29]. In α-Bi$_4$Br$_4$, this effect is reinforced by strong mass anisotropy in the conduction band ($m_{e,a} \approx 2.9m_0$ and $m_{e,b} \approx 0.09m_0$, where $m_0$ is the free electron mass), which suppresses diffusion away from the chain direction, promotes charge accumulation, and amplifies the dynamic gap reduction. A schematic of the relaxation pathways (Fig. 3e) illustrates intraband scattering and bulk-to-edge transitions, with the latter (red) suppressed due to limited phase space. Notably, photocarriers (black) are mostly localized along the *x*-axis, giving rise to the polarization sensitivity shown above. Figure. 3d also shows that at high carrier densities ($n_e$>2.16×10$^{20}$ cm$^{-3}$), the recovery time of the bandgap at 2 eV is nearly twice the carrier lifetime, whereas at 1.5 eV it remains comparable to both ESA/BF lifetimes and the hot-electron relaxation time obtained from time-resolved photoemission Fig. S7b).

Quantitative understanding of the light-induced band modulation can be obtained from doping-dependent DFT calculations. Figure 4a presents the calculated band structure of α-Bi$_4$Br$_4$ along the Γ–Z–L–M directions (inset), with orbital-resolved transition probabilities (*P*) for *b*-polarized absorption in the 1.9–2.1 eV range overlaid as color-coded markers; the marker size denotes transition magnitude, red–blue gradient distinguishes $p_x$/$p_z$ from $p_y$ orbitals (Figs. S8 and S9). The transient reflectance feature near ~1.9 eV originates mainly from VB2 to CB6 transitions at ~1.7 eV, while contributions from other k-paths such as Γ–X are negligible. Comparison of total transition probabilities (Fig. 4b) shows that *b*-polarized absorption along Γ–Z is nearly twenty times stronger than *a*-polarized, consistent with the anisotropy observed experimentally (Fig. 2a). In addition, the heavy effective masses calculated along Z confirm strong carrier localization, which enhances Coulomb screening and exchange–correlation effects, thereby amplifying the band modulation.

Upon electron doping, both VB2 and CB6 shift downward nonlinearly with carrier density, as shown by the Γ-point energies in Fig. 4c over the same $n_e$ range as the experiments. However, their separation decreases nearly linearly with $n_e$, yielding a quasi-linear modulation of the bandgap and associated absorption resonance (Fig. 4d). This deviates from the quadratic dependence of the optical gap renormalization observed experimentally (Fig. 2c), although the absolute magnitudes converge at high densities ($n_e > 1 \times 10^{20}$ cm$^{-3}$). The mismatch at lower densities may reflect underestimated shifts in the calculations, neglected correlation effects, or additional contributions from coherent phonon–driven modulation.

Mode analysis identifies the two coherent oscillations in the transient spectra as A$_g$ shearing modes with atomic displacements perpendicular to the chains (Fig. 4e) [16]. These vibrations modulate electronic polarization along the same directions, predominantly affecting the $p_x$/$p_z$ orbitals, as confirmed by the differential charge distributions (blue and yellow isosurfaces). This directly influences the unoccupied states involved in the VB2 to CB6 optical transition near 2 eV. Figure 4f quantifies the bandgap modulation versus normal mode coordinates, revealing shifts of ~1 meV that agree with experiment. The combined effect of both modes (dashed curves in Fig. 4e) illustrates the extrema of bandgap tuning (black for maximum, green for minimum) emerging at different delays as the phonons evolve.

We emphasize that light-driven modulation of electronic bands arises from the combined influence of electronic and lattice degrees of freedom. Among these, the coherent phononic contribution—often subtle in many materials—should be regarded as a general mechanism for tuning material properties with ultrahigh precision. In α-$Bi_4Br_4$, this effect is strongly manifested, motivating a minimal tight-binding description of the Bi zigzag chain, where the bands near the probe energy are dominated $p_x$ and $p_z$ orbitals. The lattice (Fig. S10) is parameterized by the intrachain distance $2h$ with hopping strength $q$, and the interchain spacing $l$ with hopping $p$ [33]; generalization to isotropic systems can be achieved by adjusting the hopping strengths. The Hamiltonian is thus:

$$H = -\sum_i \psi_i^\dagger \hat{T} \psi_{i+1} + h.c., \qquad (1)$$

where $\psi_i = (a_i, b_i)^T$ is the two-component spinor for the sublattice, and the hopping matrix is given by $\hat{T} = \begin{pmatrix} q & p \\ p & q \end{pmatrix}$, and the on-site interactions are neglected for simplicity. A 3D view of the modeled band structure is shown in Fig. 3e. The strong anisotropy between $p$ and $q$ creates a momentum-space imbalance of electron populations, leading to quasi-1D carrier accumulation and enhanced screening under photoexcitation.

Based on the two-band model, we construct a framework that incorporates both incoherent carrier screening and coherent phonon modulation of the band structure. The evolution of the gap can be expressed as:

$$\Delta E \propto -A n_e^2 e^{-\frac{t}{\tau_{BR}}} + 4p_0 \left[ 1 - \sum_j A_{ph,j} \cos(2\pi \omega_j t) e^{-\frac{t}{\tau_{ph,j}}} \right], \qquad (2)$$

where the first term phenomenologically describes a quadratic density-dependent reduction of the gap with lifetime $\tau_{BR}$, and the second term accounts for mode-selective oscillations driven by coherent phonons of frequency $\omega_j$ and lifetime $\tau_{ph,j}$ (See Methods for details). Modulating the hopping strength by electrons could also arise from a Floquet-like mechanism via a Peierls substitution [40, 48], but it usually occurs on the timescale of the pulse width. Consequently, the carrier contribution produces a monotonic narrowing of the bandgap as carriers relax, while the phonon contribution imprints periodic modulations that decay on their respective lifetimes. The superposition of these two effects governs the transient optical response, in excellent agreement with the spectra and carrier-density–dependent gap shifts observed experimentally (Fig. S10).

Finally, because the bandgap can be dynamically modulated by coherent phonons, we propose and demonstrate a general strategy for active, mode-selective band engineering using coherent control techniques [49, 50]. In a dual-pump scheme, the inter-pump delay $\Delta t_p$ is scanned over one cycle of the 0.68 THz phonon, while the probe monitors transient reflectivity for two phonon periods to capture the gradual evolution of the coherent phonon amplitude (Fig. 5a). By adjusting $\Delta t_p$, the phonons are driven in between in phase and out of phase configurations, leading to a continuous tuning between twice the enhancement and nearly complete suppression of the excitation; two extrema transient reflectivity curves are shown in Fig. 5b This control allows continuous tuning of the bandgap on the picosecond scale, achieving a precision as fine as ~100 μeV in our experiment, with a dynamic range up to ~1.5 meV (Fig. 5c). We note that the ultimate accuracy for such tuning is only capped by the thermodynamic limit of the corresponding phonon population. In this way, the band energy can be positioned at will by adjusting $\Delta t_p$, offering precision access to subtle quantum phases. These results demonstrate that coherent phonon excitation, whether through a Raman mechanism as in our result, or via resonant absorption using long-wavelength stimuli, serve as a powerful knob for ultrafast band-structure engineering and deterministic, mode-resolved control of electronic properties in quantum materials.

In summary, we have achieved joint incoherent and coherent tuning of the electronic band structure in the quasi-1D topological insulator α-Bi$_4$Br$_4$ under femtosecond optical excitation. While carrier-driven screening induces a nonlinear narrowing of the optical gap, coherent phonons generate mode-selective oscillatory modulations. A minimal tight-binding framework captures these mechanisms and, together with a dual-pump control scheme, establishes a general all-optical strategy for steering electronic bands on the intrinsic timescales of lattice vibrations. This approach enables ultrafast band engineering with unprecedented μeV-level precision. Looking ahead, extending this framework into the nonlinear regime could allow strong-field excitations to drive electron- and phonon-assisted Floquet dressing [15, 36, 51] permitting intertwined control over both hopping amplitudes and phases. Moreover, mapping time-domain modulation into real space via propagating coherent phonons may provide delocalized, pulsed control of band structures, opening pathways toward all-optical optoelectronic and optothermal devices [52, 53], as well as phonon-based quantum transducers and processors [54-56]. Our results thus offer a versatile methodology for dynamical band-structure engineering, advancing both the fundamental understanding of light–matter interactions and the practical foundations of ultrafast modulation of electronic structures from solids to molecular and biological materials.

## Methods:

**Transient reflectivity and reflectance spectroscopy.**

The femtosecond transient reflectance spectroscopy setup employs a wavelength tunable optical parametric amplifier (OPA, ORPHEUS-HP) for the generation of the pump pulse (700-950 nm), and a broadband white light supercontinuum as the probe pulse (650–1000 nm). Both pump and probe are operated at 100 kHz, the pump pulse is chopped at 4 Hz and 513 Hz for transient reflectance and transient reflectivity measurements. The pump and probe pulses are combined to co-propagate and directed through a 50X objective lens (MY50X-805) to be normally incident on the $Bi_4Br_4$ sample placed in a Montana cryostat (Montana Instruments). The sizes of the pump and probe spots are both ~2 μm in diameter. The pump pulse duration was measured to be ~80 fs at the sample. The reflected probe was either sent to a silicon photodiode and measured via a lock-in amplifier (Stanford SR830) in the case of a narrow-band transient reflectivity measurement, or to a spectrometer (HORIBA, iHR 550) in the case of transient reflectance spectra measurement. All measurements were performed with the sample temperature held at 13 K.

**Density functional calculations.**

The density functional theory (DFT) [57] calculations were implemented by the Vienna Ab Initio Simulation Package (VASP) [58] using the projector-augmented plane wave (PAW) approach [59]. The exchange correlation functional is described within the generalized gradient approximation (GGA) of Perdew-Burke-Ernzerhof (PBE) version [60]. The energy cutoff was set 400 eV and a Γ-centered k-point grid of 7×7×2 was adopted for the sampling within the Brillouin zone. The van der Waals interactions were correlated by the DFT-D3 method with Becke-Johnson damping function [61]. The convergence criteria for total energies and forces were set to $1\times10^{-10}$ eV and $1\times10^{-5}$ eV/Å, respectively. The phonon dispersions and phonon modes were calculated by density functional perturbation theory (DFPT) approach [62] without spin-orbit-coupling (SOC) using the PHONOPY package [63]. All other calculations include the SOC effect.

**Steady state optical and Raman spectroscopies**

Polarized Raman characterization of thin flake α-$Bi_4Br_4$ samples was taken at temperature of 1.7 K in an attodry2100 cryostation. The samples were exfoliated on $SiO_2$ base chips in air right before inserting into the cryogenic environment. A He-Ne laser of wavelength 633 nm was used for illumination, with a typical power of 0.5 mW, and beam size of 3 μm in diameter. Measurements were repeated several times for data consistency.

Polarized optical spectroscopy measurements were conducted at room temperature for thin flake materials exfoliated on sapphire substrates. A xenon lamp light source was used to illuminate the sample with the detection area of 25×25 μm² area. Polarized detection was achieved by inserting linear polarizers in the beam path, and the polarization directions was chosen to be parallel or perpendicular to the distinct exfoliation edges of the sample.

**TR-PEEM experiment**

The NIR-UV two color time-resolved micron-area photoelectron spectra was acquired in a commercial PEEM II apparatus (PEEMIII, Elmitec). The excitation femtosecond light source for photoemission measurements was provided by a home-built noncollinear optical parametric amplifier (NOPA) system, pumped by a high-power Yb-fiber laser (YF-FL-50-200-IR). The NOPA typically delivers pump pulses with energies of 100–500 nJ, durations of ~15 fs, and tunable wavelengths spanning the visible to near-infrared range. The UV probe pulse is generated from the 4-th harmonic generation of the NIR pump pulse without further compression. For the measurement in Fig. S7, the pump and probe pulses were selected to have energies of 1.5 eV and 6 eV, respectively. Their cross correlation is of ~100 fs. Both the pump and probe pulses are directed colinearly into the PEEM chamber at normal incidence for the transient photoelectron spectroscopy measurements. The photoelectron spectroscopy mode operates in the dispersive mode of the PEEM III.

## Evaluation of phonon-driven band renormalization

The frequency components of the time-domain band renormalization signal $\Delta E(t)$, after subtracting the incoherent background, were obtained using Fourier analysis:

$$Y(\omega_k) = \sum_{n=0}^{N-1} \Delta E(n\Delta t) e^{-i2\pi\omega_k n\Delta t} \qquad (4)$$

where $\Delta t = 0.1$ ps is the sampling interval, $N$ is the number of data points, $n$ is the time index corresponding to the discrete sampling time $\omega_k = k/(N\Delta t)$ denotes the discrete frequency corresponding to frequency bin index k=0, 1, …, $N-1$. The Fourier amplitude (FA) spectrum at frequency $\omega_k$ was calculated as:

$$A_{FA}(\omega_k) = \frac{2}{N}|Y(\omega_k)| \qquad (5)$$

which yields the real magnitude amplitude of the oscillation at frequency $\omega_k$. The discrete Fourier frequency nearest to each phonon mode at 0.68 and 1.26 THz was selected for the corresponding assignment This gives two oscillating modes of frequencies $\omega_{k1}$=0.69 and $\omega_{k2}$=1.26 THz, with an amplitude of $A_{FA}(\omega_{k1}) \approx 1.44$ meV and $A_{FA}(\omega_{k2}) \approx 1.12$ meV, respectively.

A direct fit with trigonometric functions was also performed:

$$\Delta E(t) = \sum_{j=1}^{2} A(\omega_{\text{fit},j}) \cos\left(2\pi(\omega_{\text{fit},j} t + \varphi_{\text{fit},j})\right) \qquad (6)$$

This yields two modes having frequencies $\omega_{\text{fit},1}$=0.81 ± 0.01 and $\omega_{\text{fit},2}$=1.26 ± 0.01 THz, with amplitudes of $A(\omega_{\text{fit},1})$=1.43 ± 0.18 meV and $A(\omega_{\text{fit},2})$=1.31 ± 0.17 meV, respectively.

**Acknowledgments**
Computational time was supported by the Center for Computational Science and Engineering of Southern University of Science and Technology and the Major Science and Technology Infrastructure Project of Material Genome Big-science Facilities Platform supported by Municipal Development and Reform Commission of Shenzhen.
**Funding**
This work was supported by the Ministry of Science and Technology of China (2024YFA1409800, 2022YFA1402903 and 2024YFA1409101), the National Natural Science Foundation of China (12374223, 92477108, and 12374059), the Shenzhen Science and Technology Program (20231117151322001 and JCYJ20240813095301003), and the Guangdong Provincial Quantum Science Strategic Initiative, the National Key Research and Development Program of China (2020YFA0308800 and 2022YFA1403400), the Beijing Natural Science Foundation (Z210006), and the Beijing National Laboratory for Condensed Matter Physics (2023BNLCMPKF007)

**Author contributions:** Y.N.D. and L.H. supervised the work. Y.Q.Z. performed crystal growth under the supervision of Z.W.W. P.T. performed the transient reflectance measurements and data analysis, with the help of M.H.S. Y.T.C. performed the DFT calculations. H.Y.C. performed the polarized Raman experiments under the supervision of M.Y.H. M.C. and T.P. performed the steady-state optical spectroscopy experiments under supervision of Y.J.L. G.Y.X. performed the time-resolved PEEM measurements with the help of J.X.Y. P.T. and Y.N.D. wrote the manuscript. All authors discussed the results and commented on the manuscript.
**Competing interests:** Authors declare that they have no competing interests.
**Data and materials availability:** The data that support the reported findings of this study are available from the corresponding authors on reasonable request.

**Figure 1**

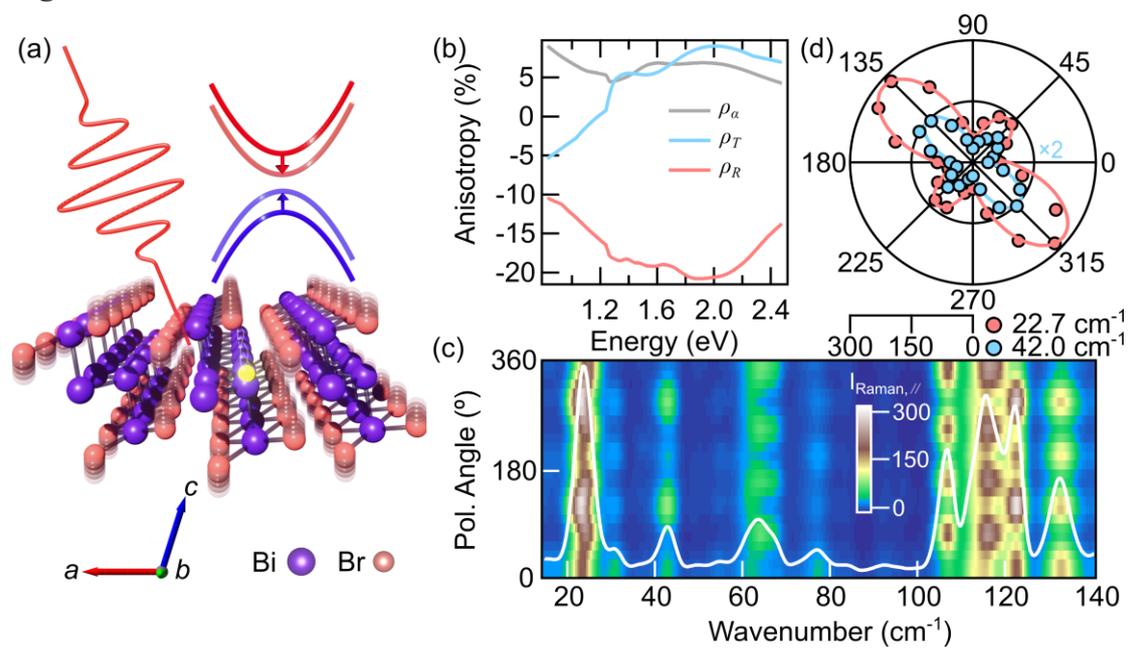

**Figure 1.** Schematic of band engineering and optical characterization. (a) Schematic of band engineering induced by photoexcited quasiparticles. The crystal structure of α-Bi$_4$Br$_4$ is shown, with Bi (purple) and Br (red) atoms forming quasi-1D chains along the *b*-axis. Band modification arises from both charge carriers (yellow sphere) confined along the chains and coherent phonon motions perpendicular to them. (b) Anisotropy ratios of absorption $\rho_\alpha$ (gray), transmission $\rho_T$ (blue) and reflection $\rho_R$ (red) as a function of the incident light energy. (c) Polarization-resolved Raman pseudocolor map acquired in the parallel (//) polarization configurations. The polarization angle is defined such that 0° corresponds to the *b*-axis (chain direction) and 90° to the *a*-axis (perpendicular) of the crystal. The white curve represents the polarization integrated Raman spectrum for quantitative mode assignment. (d) The Raman intensity in polar coordinates for modes at 22.7 cm$^{-1}$ and 42.0 cm$^{-1}$ (markers) along with the corresponding fitted angular dependences (curves).

Figure 2

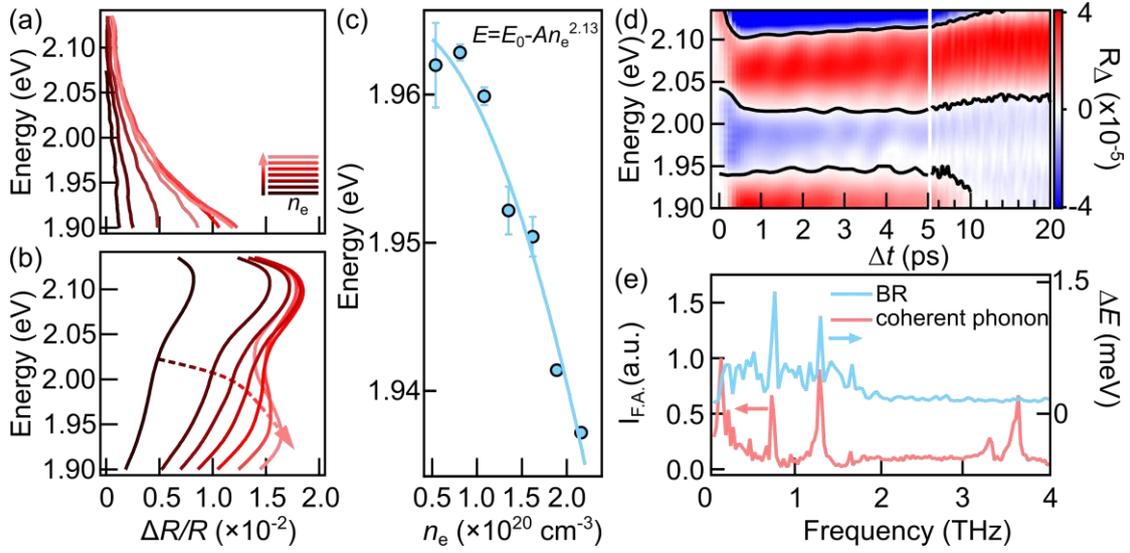

**Figure 2.** Coherent and incoherent band engineering of α-Bi4Br4. (a) Initial transient reflectance spectra at 13 K for carrier densities $n_e$=0.54-2.16×10$^{20}$ cm$^{-3}$ (black to red). Pump polarization is along the *b*-axis and probe polarization along the *a*-axis. (b) Same as (a) but with probe polarization along the *b*-axis. Dashed arrows indicate peak shifts. (c) Resonance energies extracted by Lorentzian fits to (b), overlaid with a power-law fit $E = E_0 - An_e^\beta$, where A is a proportionality factor, and $E_0$ is the intrinsic energy without photoexcitation. (d) Differential transient reflectance spectra $R_\Delta(\omega)$ at $n_e$ =2.16×10$^{20}$ cm$^{-3}$. White color represents the zero point $R_{\Delta,0}(t)$. Black lines trace the temporal evolution of spectral extrema near 1.95, 2.0, and 2.1 eV as in (a). (e) Normalized Fourier spectrum of the band renormalization (BR, right axis) at 2.0 eV from (d), compared with coherent phonon modes (left axis) extracted from Fig. S3.

Figure 3

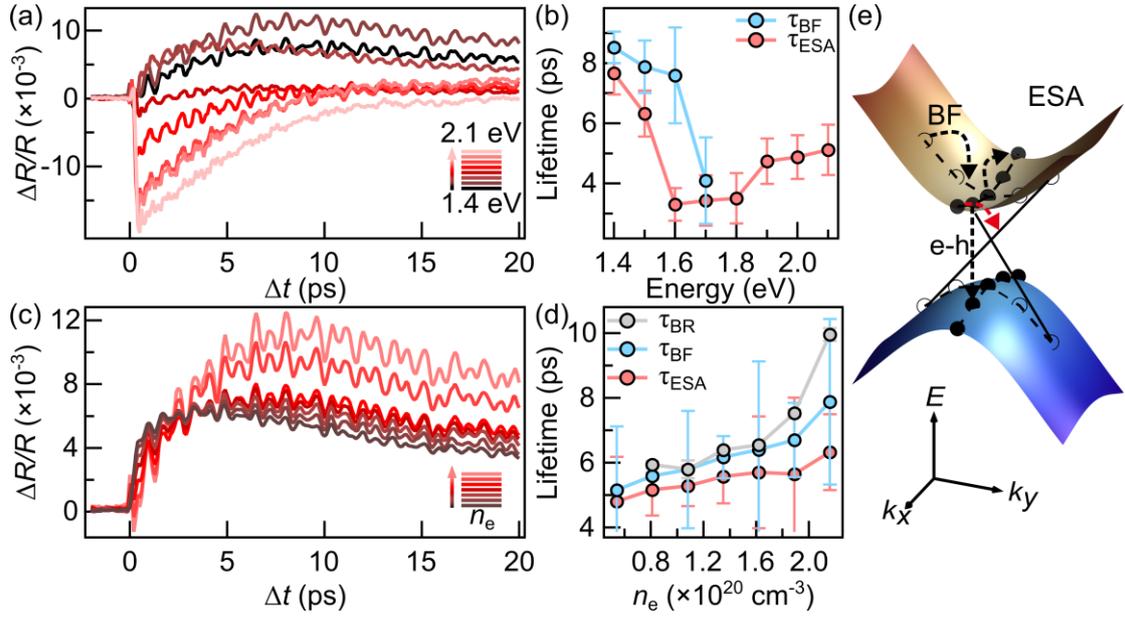

**Figure 3.** Mechanisms for carrier-driven band modification in α-Bi4Br4. (a) Probe energy dependent differential reflectivity (*ΔR/R*) spectra, with color scale from black to pink indicating increasing probe energy. (b) Energy-dependent lifetimes extracted from (a). Blue denotes BF and red ESA; BF is mostly pronounced only for energies ⩽ 1.7 eV. (c) Carrier-density–dependent *ΔR/R* spectra, with color scale from black to red indicating increasing $n_e$ from $0.54 \times 10^{20}$ cm$^{-3}$ to $2.16 \times 10^{20}$ cm$^{-3}$. (d) Energy-dependent lifetimes extracted from (c). Blue and red circles denote BF and ESA, respectively, while yellow circles represent BR lifetimes obtained by fitting the zero-crossing (black curve) near 2.1 eV in Fig. 2d. Negligible BR with large variance was obtained at the lowest carrier density. (e) Schematic of anisotropic band structure and carrier dynamics. Solid black spheres represent charge accumulation, hollow spheres unoccupied states. Anisotropic dispersion leads to carrier localization along $k_x$ but delocalization along $k_y$. Dashed arrows indicate relaxation channels, with the red arrow suppressed due to limited phase-space scattering.

**Figure 4**

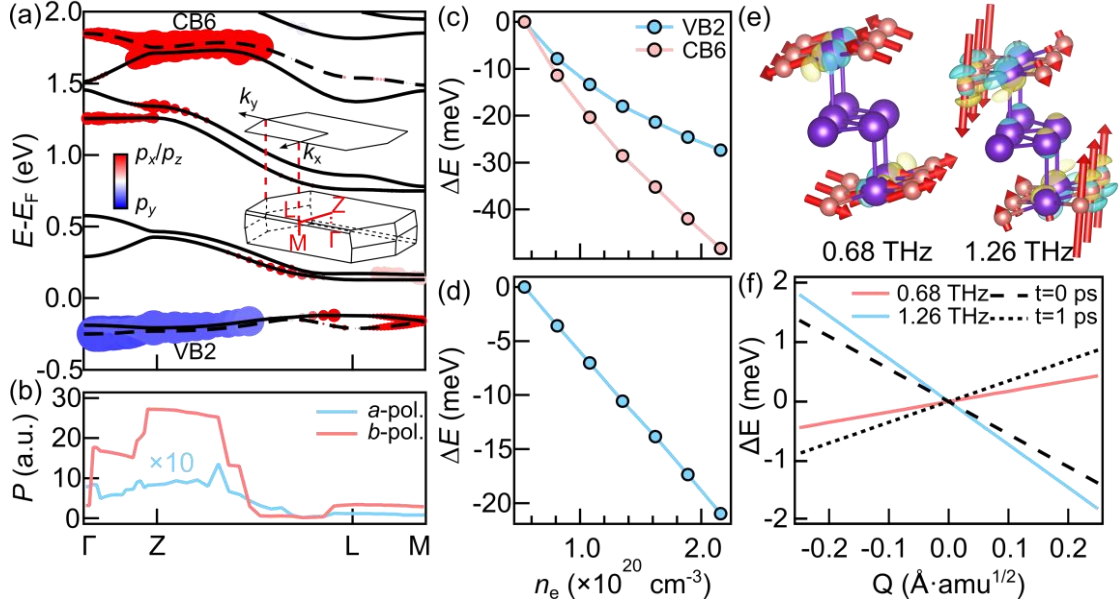

**Figure 4.** First principles calculations on the renormalization effects of α-Bi$_4$Br$_4$. (a) Calculated electronic band structure with superimposed markers indicating orbital composition (blue–white–red gradient) and transition probability $P$ (marker size) integrated in the 1.9–2.1 eV range. Both ground-state and excited-state transitions are included. Inset: Brillouin zone of bulk α-Bi$_4$Br$_4$. (b) Polarization-resolved transition probabilities $P$ in (a): blue for $a$-axis polarization, red for $b$-axis polarization. (c) Energy shifts at Γ for VB2 (blue) and CB6 (red), as indicated by dashed bands in (a), plotted versus carrier density $n_e$. (d) Bandgap between VB2 and CB6, derived from (c), as a function of $n_e$. (e) Differential charge distributions for the 0.68 THz and 1.26 THz phonon modes. Yellow and blue isosurfaces represent charge accumulation and depletion, respectively; red arrows on atoms indicate displacement directions and magnitudes. (f) Bandgap modulation as a function of phonon normal mode coordinate $Q$. Red and blue lines correspond to the 0.68 THz and 1.26 THz modes, respectively. Dashed lines show their combined effect at 0 ps and 1 ps, corresponding to maximum and minimum band tuning. In-phase excitation with identical $Q$ is assumed.

**Figure 5**

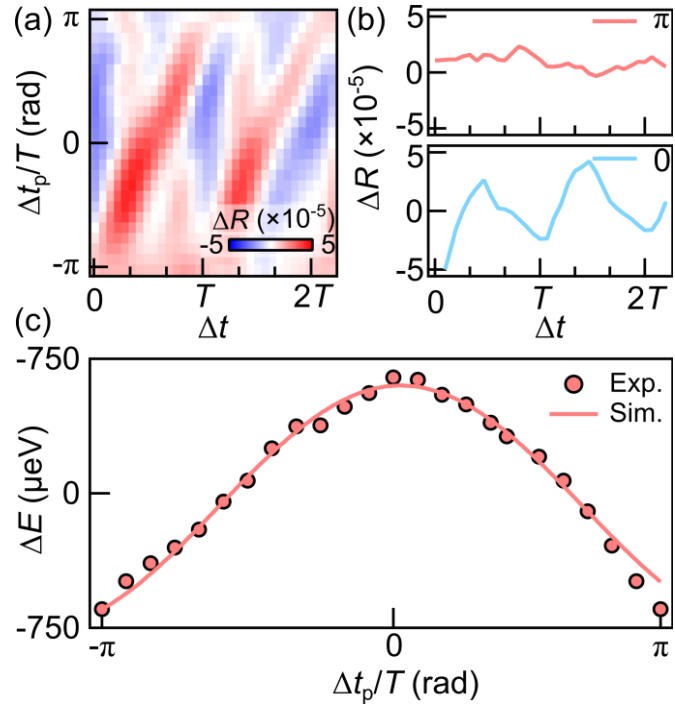

**Figure 5.** Coherent control of the optical gap with μeV precision. (a) Differential reflectivity as a function of pump-pump delay per phonon period (0.68 THz), $\Delta t_p/T$ (in unit of phonon phase) and the pump-probe delay $\Delta t$ (in unit of $T$), after subtraction of the incoherent background. $\Delta t$ is defined as the probe delay relative to the first pump. (b) Line profiles from (a) at $\Delta t_p = \pi$ (red, out-of-phase, $A_{1g}$ mode suppressed), and at $\Delta t_p = 0\pi$ (blue, in-phase, $A_{1g}$ mode enhanced). (c) $\Delta t_p$-dependent energy modulation by the 0.68 THz phonon with the corresponding tight-binding model simulation.